# Spin-orbit interactions and chiroptical effects engaging orbital angular momentum of twisted light in chiral and achiral media

Kayn A. Forbes and David L. Andrews

*School of Chemistry, University of East Anglia, Norwich, NR4 7TJ, UK*

**ABSTRACT**

There is recurrent interest in the orbital angular momentum (OAM) conveyed by optical vortices, which are structured beams with a helically twisted wavefront. Particular significance is attached to the issue of how, in its interactions with matter, light conveying OAM might prove sensitive to the relative handedness and degree of twist in the associated optical wavefront. As a result of recent experimental and theoretical studies, the supposition that beams with OAM might enable discrimination between oppositely handed forms of matter has become a renewed focus of attention. Some of the tantalizing conclusions that are beginning to emerge from this research have, however, not yet established a definitive basis for a supporting mechanism. To resolve this problem requires the development of theory to support a faithful representation, and a thorough understanding, of the fundamental molecule-photon physics at play in such optical processes – even for processes as basic as absorption. The present analysis establishes mechanisms at play that entail a novel manifestation of optical spin-orbit interactions (SOI), engaging transition electric quadrupole moments. Moreover, powerful symmetry principles prove to render distinctively different criteria governing the exhibition of 2D and 3D chirality. The new results elucidate the operation of such effects, identifying their responsibility for discriminatory optical interactions of various forms in both chiral and achiral media.

## I. INTRODUCTION

The primary character of an optical vortex, or twisted light, is its chiral form. This is a feature throughout the field of singular optics, since a singularity in any waveform is invariably associated with a local phase ramp [1] – and as light has a propagating character, 3D helicity must inevitably result. It has therefore long been a topic of interest to speculate on whether, or by what means, twisted light might engage in a characteristic way with a physical object or system that is chiral. To exhibit chirality, both optical radiation and matter are subject to the same criterion: its key properties cannot be eigenstates of the space inversion operation. The context for such a question is, of course, a world subject to the rule of symmetry through conservation laws [2]. In the specific field of optical interactions between light and molecules, an interplay of material and optical chirality has already been seen to play an extremely important role in many phenomena. In general, chiroptical mechanisms interlace the chirality of molecules with the helicity of light associated with polarization, to exhibit process rates and forces that are sensitive to changes of either optical or material handedness in the total system [3]. Classic examples include circular dichroism [4], differential Rayleigh and Raman scattering [5], resonance energy transfer and discriminatory dispersion interactions [6,7], and optical trapping and binding forces [8,9]. The latter examples are relatively recent discoveries, which show a propensity for the optical separation of enantiomers and identification of the actual existence of chirality in a system [10].

While polarization is a feature that can be controlled and exhibited by all forms of optical radiation, the burgeoning field of structured light presents opportunities to explore another dimension of chirality. In particular, work on the development of twisted light has led to the striking identification and observation that such beams, and indeed the single photons they comprise, can possesses an intrinsic optical orbital angular momentum (OAM) [11–13]. While an intrinsic photon spin angular momentum (SAM) of $\pm\hbar$, is manifest in left- and right-handed circularly polarizations, crucial for most of the chiroptical interactions mentioned above, light possessing a vortex structure – with an intensity singularity at the center of the beam – exhibits an orbital angular momentum per photon of $\pm\ell\hbar$, where the *topological charge* $\ell$ may take any integer value. The topological charge, or *winding number*, signifies the multiplicity (within a wavelength) and direction of twists in the phase-front. Evidently, such structured forms of light possess handedness in a similar way to circularly polarized light: the handedness of a vortex beam is due to the helical form of the wave-front for the propagating beam; it has nothing to do with the polarization degree of freedom, whereas the handedness of a circularly polarized beam is due to the helix that the electric (and magnetic) field vector traces out.

As such, it may legitimately be asked, just as the handedness of circularly polarized light produces chiroptical interactions, whether the handedness of a twisted beam can also produce discriminatory optical processes with matter. In other words, can any piece of matter interact differently with a right-handed vortex beam than with a left-handed one? An initial answer to such a question was first presented nearly two decades ago, using quantum electrodynamical theory (QED) to reach a conclusion that the handedness of a structured light beam can play no role in chiroptical interactions [14]. Provoked by the theoretical work, complementary and supportive experimental observations that soon followed satisfied the experimental conditions assumed by the theory in its derivation [15,16]. However, in the last few years, further experimental studies, looking at systems under different conditions, have in contrast appeared to show the contrary – the handedness of twisted light *can* exert a chiroptical influence. These studies, invoking spin-orbit interactions (SOI) [17] of the incident light, have been able to induce chiroptical effects with OAM by utilizing the helicity-dependent intensity distributions that occur due to the SOI of focused non-paraxial vortex beams with circular polarization. Broadening the definition of 'circular dichroism', fundamentally related effects have been identified in non-chiral nanostructures [18]; furthermore, effects of a similar kind have been discovered in achiral atomic matter [19] and in the use of so-called spin-orbit beams to characterize material chirality [20]. Other studies have investigated the exploitation of plasmonic coupling in light-matter interactions with twisted light to engineer chiroptical effects [21–25], with similarly dichroic effects manifest with vortex electron beams [26], atomic Bose-Einstein condensates [27] and OAM-induced X-ray dichroism [28]. There has even been experimental work looking at so-called 'magneto-orbital' dichroism, an OAM analogue of magnetic circular dichroism [29], along with theoretical work looking at the use of stimulated parametric down conversion to produce a dichroic-like effect through the direct action of the OAM of an incident field [28].

To date, relatively few studies describing the spin-orbit interactions (SOI) of light appear to have concerned freely propagating paraxial light [31,32]; most such studies have involved non-paraxial optical fields (as in focused or scattered light), focusing upon effects in inhomogeneous media and at interfaces and metasurfaces, as well as through the engagement of evanescent near-fields [17,33]. These SOI interactions lead to *spin-to-orbit AM conversion* of light, as well as the *spin Hall* and *orbital Hall effect* of light. Moreover, most of theory, in both the classical and quantum regime, have been restricted to the dipole approximation.

The rates of single-photon absorption by vortex photons derived in this paper are distinctly different in that, in their theoretical derivation, each photon in the beam is assumed to be freely propagating in a paraxial fashion, within the Rayleigh range. None of the SOI addressed above are specifically applicable to this situation: the effects that are now being identified relate to the direct interaction of the light with individual particles of matter, without the beam being subjected to any scattering or focusing, for example. We now significantly build upon the most recent exploratory theoretical work, where a more involved theory of QED has highlighted a discriminative mechanism at play in single-photon absorption for both chiral and achiral molecules, in which the key roles of quadrupole interactions and molecular orientational effects have been identified [19,34,35]. These are issues that were overlooked in previous studies [14]. Accounting for the specific optical process of circular dichroism, the new theory has, however, created space for a host of supplementary questions; moreover, in its present form it has given relatively little insight on principles of wider generality. The analysis left unanswered some questions, such as why it appeared necessary for the discriminating twisted light beam to also possess a circular state of polarization. It is therefore timely to progress from a relatively simple investigation of how one can produce these OAM-sensitive chiroptical interactions, to securing the underlying mechanisms, and explaining why and how they work. In this paper we take steps towards elucidating the latent physics, recognizing that the interactions are explicit manifestations of spin-orbit interactions in the light. Furthermore, by carefully considering the underlying symmetry principles, we identify completely novel results associated with 2D chirality – in other words, effects exhibiting a sensitivity to the sense of wavefront twist, in material systems that do not conform to the usual criteria for the chirality of 3D structures.

In Section II we begin with an outline of the general QED theory for matter interacting with the most commonly employed form of vortex light: Laguerre-Gaussian (LG) beams. The origins of chiroptical effects with optical OAM are explicitly highlighted with regards to the electric field gradient and electric quadrupole transition moments; Section III then employs the theoretical foundations laid out in II to derive the rate of single-photon absorption of LG photons in chiral matter. Section IV extends the theory to account for chiroptical interactions with vortex light in achiral matter; Section V then highlights the fact that the chiroptical processes that have been detailed are actually a manifestation of a form of SOI that takes place during the electronic transitions when the photons are absorbed. Finally, in Section VI, it is shown that certain chiroptical effects that fall out of the theory are dependent on *only* either the topological charge of the vortex beam *or* the circular polarization state, and are thus not SOI (in contrast to the phenomena derived in the previous sections); the role that 2D chirality plays in these exotic interactions is then explained. We conclude by highlighting potential routes to further this exciting field of research in the immediate future.

## II. QUANTUM ELECTRODYNAMICS AND TWISTED LIGHT

The multipolar Hamiltonian is best used to study the electrodynamics of light-matter interactions, and when it is expressed in fully quantized form, as in QED, it is most commonly referred to as the Power-Zienau-Woolley (PZW) Hamiltonian [36]. This gauge invariant PZW Hamiltonian has clear advantages in non-covariant analyses (generally applicable where it is not necessary for electron motions to be corrected for relativistic effects), and has recently even provided profound application in condensed matter systems [37–40]. The molecular electric and magnetic molecular multipole moments couple directly to the physically realizable electric and magnetic fields:

$$H_{\text{int}} = \sum_{\xi}\left[-\boldsymbol{\mu}(\xi)\cdot\boldsymbol{e}^{\perp}(\boldsymbol{R}_{\xi}) - Q_{ij}(\xi)\nabla_j e_i^{\perp}(\boldsymbol{R}_{\xi}) - \boldsymbol{m}(\xi)\cdot\boldsymbol{b}(\boldsymbol{R}_{\xi})... + h.o.t.\right] + \frac{e^2}{8m}\sum_{\xi,\alpha}\left[\left(\boldsymbol{q}_{\alpha}(\xi) - \boldsymbol{R}_{\xi}\right)\times\boldsymbol{b}(\boldsymbol{R}_{\xi})\right]^2 ... + ...h.o.t., \quad (1)$$

where for a molecule $\xi$ positioned at $\boldsymbol{R}_{\xi}$, $\boldsymbol{\mu}$ is the transition electric dipole (E1) moment operator; $\boldsymbol{Q}$ is the transition electric quadrupole (E2) operator and $\boldsymbol{m}$ is the transition magnetic dipole (M1) moment operator; the final term in (1) is the leading order diamagnetic interaction term [41,42], and $\boldsymbol{q}_{\alpha}(\xi)$ is the position vector of an electron $\alpha$ possessing a charge $e$ and mass $m$. The first term in (1) thus represents E1 coupling, the second E2 and the third M1; $\boldsymbol{e}^{\perp}(\boldsymbol{R}_{\xi})$ is the electric field and $\boldsymbol{b}(\boldsymbol{R}_{\xi})$ is the magnetic field. It is worth highlighting at this stage that the superscript ⊥ designates that, in this electrodynamic gauge, the electric field is quantized in terms of transverse photons (the magnetic field being necessarily transverse in any gauge): this transversality is not to be identified with the completely different sense discussed in later sections concerned with transverse and longitudinal field components, relative to the propagation direction of the beam. The electric and magnetic field vacuum mode expansions for Laguerre-Gaussian beams, in the paraxial approximation, emerge as functions of the cylindrical coordinates [43,44]: the off-axis radial distance $r$, axial position $z$ and azimuthal angle $\phi$;

$$e^{\perp}(r) = i \sum_{k,\eta,\ell,p} \left( \frac{\hbar c k}{2\varepsilon_0 V} \right)^{1/2} \left[ e_{\ell,p}^{(\eta)}(k) a_{\ell,p}^{(\eta)}(k) f_{\ell,p}(r) e^{(ikz+i\ell\phi)} - \bar{e}_{\ell,p}^{(\eta)}(k) a_{\ell,p}^{\dagger(\eta)}(k) \bar{f}_{\ell,p}(r) e^{-(ikz+i\ell\phi)} \right], \qquad (2)$$

and

$$b(r) = i \sum_{k,\eta,\ell,p} \left( \frac{\hbar k}{2\varepsilon_0 c V} \right)^{1/2} \left[ b_{\ell,p}^{(\eta)}(k) a_{\ell,p}^{(\eta)}(k) f_{\ell,p}(r) e^{(ikz+i\ell\phi)} - \bar{b}_{\ell,p}^{(\eta)}(k) a_{\ell,p}^{\dagger(\eta)}(k) \bar{f}_{\ell,p}(r) e^{-(ikz+i\ell\phi)} \right], \qquad (3)$$

where $a_{\ell,p}^{(\eta)}(k)$ and $a_{\ell,p}^{\dagger(\eta)}(k)$ are the annihilation and creation operators for a photon of mode $(k,\eta,\ell,p)$; $e_{\ell,p}^{(\eta)}(k)$ and $b_{\ell,p}^{(\eta)}(k)$ are the unit polarization vectors transverse to $k$, such that $b_{\ell,p}^{(\eta)}(k) = \hat{k} \times e_{\ell,p}^{(\eta)}(k)$; and, for a beam of waist $w_0$, the radial distribution function $f_{\ell,p}(r)$ is

$$f_{\ell,p}(r) = \frac{C_p^{|\ell|}}{w_0} \left[ \frac{\sqrt{2}r}{w_0} \right]^{|\ell|} e^{(-r^2/w_0^2)} L_p^{|\ell|}\left( \frac{2r^2}{w_0^2} \right). \qquad (4)$$

In the above (4), $C_p^{|\ell|}$ is simply a normalization constant, whilst $L_p^{|\ell|}$ is the generalised Laguerre polynomial of order $p$. An important point about $f_{\ell,p}(r)$ worth highlighting at this early stage is that it is only dependent on the modulus of the topological charge $|\ell|$, and not the sign of $\ell$. The most significant part of (2) and (3) is the azimuthal angular dependence contained in the term $e^{i\ell\phi}$. This azimuthal phase structure is what gives Laguerre-Gaussian modes their OAM. The term gives rise to $|\ell|$ intertwined helical wave-fronts per wavelength, and these wave-fronts can twist either clockwise or anticlockwise as they propagate. The direction of twist, determined by the sign of $\ell$, gives the helices their handedness: by common definition, for $\ell > 0$, beams twist to the left and for $\ell < 0$, to the right. In the above equations it is readily verified that when $\ell = 0$, the mode expansions reduce to a structureless Gaussian form, in the paraxial approximation. The above, along with the knowledge of standard time-dependent perturbation techniques, is all that is needed to explain and account for virtually all optical interactions with molecules [45]. For the optical processes derived in this paper, we require the following well-known and simple series to allow us to construct matrix elements for specific phenomena;

$$M_{fi}(\xi) = \langle f|H_{\text{int}}(\xi)|i\rangle + \sum_{I} \frac{\langle f|H_{\text{int}}(\xi)|I\rangle\langle I|H_{\text{int}}(\xi)|i\rangle}{E_i - E_I} + \sum_{I,II} \frac{\langle f|H_{\text{int}}(\xi)|II\rangle\langle II|H_{\text{int}}(\xi)|I\rangle\langle I|H_{\text{int}}(\xi)|i\rangle}{(E_i - E_I)(E_i - E_{II})} + \ldots \quad (5)$$

In fact, for the most simplest of optical process which we study in this work – single-photon absorption – all we require from (5) is the first term on the right-hand side, which is first-order with respect to $H_{\text{int}}$.

In a recent paper [35] we used the same QED methods outlined above to show what influence the sign of $\ell$ has on single-photon absorption with Laguerre-Gaussian beams. Such efforts led to the particularly interesting phenomenon termed circular-vortex dichroism (CVD). However, it may be argued that the most important revelation from that analysis was the apparent necessity of the transition quadrupole moments (E2) to engage the handedness of a vortex beam in *any* chiroptical interaction – as was originally shown, any combination of E1 and M1 dipole moments will never produce a chiroptical sensitivity to the OAM of the beam.

In hindsight this is rather obvious: if we wanted to observe a chiroptical sensitivity due to the sign of $\ell$, which is a representation of the phase-front, we must clearly engage a molecular multipole moment that is dependent upon the (OAM) circulating phase gradient (whilst SAM is a rotating vector property). Optical vortex beams possess both an axial field gradient and a transverse (in-plane) gradient, both of which can drive quadrupole transitions, the latter of which is manifest in part through the azimuthal phase-gradient. Evidently from (1), both the E1 and M1 have a standard linear dependence on the oscillating electric and magnetic fields, respectively. However the E2 is a function of the gradient of the dynamic electric field. Thus, the operation represented by $\nabla_j e_i^\perp$ is extremely important, and proves to pervade the associated theory:

$$\nabla_j e_i^\perp \approx \nabla_j f_{\ell,p}(r)e^{(ikz+i\ell\phi)} = f_{\ell,p}(r)\partial_r \hat{r}_j e^{(ikz+i\ell\phi)} + f_{\ell,p}(r)\frac{1}{r}\partial_\phi \hat{\phi}_j e^{(ikz+i\ell\phi)} + f_{\ell,p}(r)\partial_z \hat{z}_j e^{(ikz+i\ell\phi)}$$
$$= \hat{r}_j \partial_r f_{\ell,p}(r)e^{(ikz+i\ell\phi)} + f_{\ell,p}(r)\frac{1}{r}\left(i\ell\hat{\phi}_j - \hat{r}_j\right)e^{(ikz+i\ell\phi)} + f_{\ell,p}(r)ik\hat{z}_j e^{(ikz+i\ell\phi)}, \quad (6)$$

where we have used the shorthand notation $\partial/\partial x = \partial_x$ and the fact that $\partial_\phi \hat{\phi} = -\hat{r}$. It is immediately evident that by taking the gradient of the electric field, we produce terms linearly dependent on $\ell$ (and hence its sign); in other words, we produce terms that are dependent on the phase gradient. Therefore, the quadrupole molecular transition moments are dependent on the amount the beam is twisting, and in what direction. There is also clearly a dependence on $\ell$ in the phase factors: nonetheless, in any incoherent optical process with distinct initial and final states the observable can be cast as a rate, quadratically dependent on the matrix element when using the Fermi rate rule, and the phase factors disappear in the modulus square. [Equally, any optical phenomenon that generates energy shifts and forces will impose identical initial and final states for the total system, and thus $\Delta\ell = 0$.] At this stage, however, we cannot anticipate the general form for any chiroptical effect or form of spin-orbit interaction; these possibilities can only revealed once optical rates are calculated for specific phenomena. By including and taking account of E2 moments in an optical process, we can at least anticipate a chiroptical influence originating from the sign of $\ell$. In passing, it is worth observing that although typically small compared to the generally dominant electric dipole interaction, relatively enhanced interactions can take place between quadrupole transition moments and twisted light beams [46–49].

To proceed with calculating the optical rate of single-photon absorption using the ideas presented above we require standard time-perturbation techniques and the Fermi golden rule, such that the rate is

$$\Gamma = \frac{2\pi}{\hbar} \rho_\mathrm{f} \left| M_{fi}(\xi) \right|^2 = \frac{2\pi}{\hbar} \rho_\mathrm{f} \left| \underbrace{-\mu_i^{f0}(\xi)\langle (n-1)|e_i^\perp(\boldsymbol{r})|n\rangle}_{E1} \underbrace{-m_i^{f0}(\xi)\langle (n-1)|b_i(\boldsymbol{r})|n\rangle}_{M1} \underbrace{-Q_{ij}^{f0}(\xi)\langle (n-1)|\nabla_j e_i^\perp(\boldsymbol{r})|n\rangle}_{E2} \right|^2, \quad (7)$$

where $\rho_\mathrm{f}$ is the density of final states and $M_{fi}(\xi)$ is the quantum amplitude calculated using (5). For the final term in square brackets (the E2 term) in (7), we refer back to (6) to carry out the calculation. Evidently, through the modulus square in (7) we produce a plethora of terms: E1E1, M1M1 and E2E2, but also the interference terms E1M1, E1E2, and M1E2. As mentioned, if we wish to observe a chiroptical effect due to the OAM of the twisted photons, we must engage an E2 moment as a bare necessity – therefore, for our purposes in this article we may now concentrate on the E1E2, M1E2, and E2E2 terms.

## III. CHIRAL MEDIA

We now make an important connection between these multipole transition moments and chirality. The spatial parity signature of an E1 moment is odd, whilst that of the E2 and M1 moments is even. Here, the selection rules for electronic transitions come prominently into play. For any process involving an electronic transition, the product of the initial and final state electronic symmetries (specifically their irreducible representations known as *irreps*) must contain the symmetry irrep of one or more components of the multipole under scrutiny. For any centrosymmetric molecule it follows that, since every irrep has a definite parity, no transition can be simultaneously allowed by both E1 (odd parity) and E2 (even parity) moments; the same principle of exclusion applies to E1 and M1 simultaneity. Thus, in the case of single-photon absorption, the E1M1 and E1E2 ('µm' and 'µQ') cross-terms vanish. However, for non-centrosymmetric molecules (and indeed for chiral species in general) such terms will indeed be present in the rate equation [34]. Then, since the 'µm' and 'µQ' products have odd parity, they can be supported only by odd-parity combinations of the field vectors. On the other hand the M1E2 and E2E2 moment products can arise for both achiral (non-chiral) and chiral molecules. It is to be borne in mind that, although the E1M1 couplings have been the hallmark of the historically most widely studied forms of chiroptical interaction, their exclusion of an E2 moment makes no provision for any dependence on the handedness of twisted photons through a phase twist. We now concentrate on the E1E2 moment products to specifically entertain a situation where the material component of the system is chiral. The principles are explained in more detail in ref. [47].

All electric multipole moments are assumed real (whilst the M1 term is imaginary), and isolating the E1E2 terms therefore gives:

$$\Gamma_{\mathrm{E1E2}} = \left(\frac{n\hbar ck}{2\varepsilon_o V}\right) f_{\ell,p}^{\ 2}(r) e_i \bar{e}_k \left[ \left( \hat{r}_j \frac{1}{f_{\ell,p}(r)} \partial_r f_{\ell,p}(r) - \frac{\hat{r}_j}{r} \right) \left( \mu_i^{f0} Q_{kj}^{f0} + \mu_k^{f0} Q_{ij}^{f0} \right) + \left( i\frac{\ell}{r} \hat{\phi}_j + ik\hat{z}_j \right) \left( \mu_k^{f0} Q_{ij}^{f0} - \mu_i^{f0} Q_{kj}^{f0} \right) \right]. \quad (8)$$

For our interests here we extract the relevant terms from (8) which exhibit a dependence on the sign of $\ell$; the element of the absorption rate, for a twisted LG photon of arbitrary polarization, that depends on the sign of the topological charge by a molecule (identified in the following expression by a prime on the $\Gamma$) is thus secured as

$$\Gamma'_{\text{E1E2}}(\ell) = \frac{I(\omega)}{2c\hbar^2 \varepsilon_o} \frac{1}{r} f_{\ell,p}^{\,2}(r) e_i \bar{e}_k i\ell \hat{\phi}_j \left( \mu_k^{f0} Q_{ij}^{f0} - \mu_i^{f0} Q_{kj}^{f0} \right), \tag{9}$$

where the density of final states for the radiation has been written in terms of the irradiance per unit frequency $I(\omega)$ (which yields the intensity on integration over linewidth in frequency terms). The innocuous presence of '$i$' in (9) actually has profound physical influence upon the final result: as it stands, the right-hand side of (9) unphysically could only represent an imaginary quantity, unless the product of polarization components delivers an imaginary – or at least a complex-value – result. Thus it becomes evident that the only way to produce a real, observable rate, is to invoke circular (or at least elliptical) polarization states for the photon. Accordingly we deduce that, to the leading E1E2 order for chiral molecules, linearly polarized twisted beams will show no chiroptical effects – we require the light with an optical vortex structure to also be circularly polarized: the light must convey both an orbital and spin angular momentum along the propagation direction. The polarization-dependent factor in (9) can be simplified by using the following identity [45]:

$$e_i^{\text{L/R}} \bar{e}_k^{\text{L/R}} = \frac{1}{2} \left[ \left( \delta_{ik} - \hat{k}_i \hat{k}_k \right) \mp i \varepsilon_{ikm} \hat{k}_m \right], \tag{10}$$

which allows the rate to be expressed in terms of the helicity eigenvalues $\sigma$ for the circularly polarized state:

$$\Gamma'^{(\text{L/R})}_{\text{E1E2}}(\ell) = \sigma \ell \frac{\kappa}{r} \varepsilon_{ikm} \hat{k}_m \hat{\phi}_j \mu_k^{f0} Q_{ij}^{f0}, \tag{11}$$

where $\sigma = \pm 1$ for L or R-handed circularly polarized photons respectively; as stated previously, the topological charge can take any integer value, without upper limit. For conciseness of expression, we also now introduce the notation for the constant $\kappa = I(\omega) f_{\ell,p}^{\,2}(r) / 2c\hbar^2 \varepsilon_o$. The effect quantified by the rate (11) is clearly dependent upon both the helicity of the circularly polarized photons and the sign of the topological charge through the product $\sigma\ell$, and therefore may legitimately be considered a spin-orbit interaction. The underlying physics of the SOI in (11) is discussed in detail in Section V.

Due to the twin dependence on $\sigma\ell$ and on the molecular handedness (through $\mu_k^{f0} Q_{ij}^{f0}$), the rate (11) changes sign if two of the three different forms of handedness in the system (circular polarization state, topological charge, and molecular handedness) are fixed and the other is changed: for example, the rate is opposite in sign for a L-handed molecule absorbing a L-handed vortex photon of L-handed polarization, from that of a R-handed polarization. For a given molecular handedness, the CVD effect is therefore invariant under the transformation $(\sigma, \ell) \to (-\sigma, -\ell)$, but not $(\sigma, \ell) \to (-\sigma, \ell)$ or $(\sigma, \ell) \to (\sigma, -\ell)$.

The analysis on CVD can be made more complete by accounting for standard circular dichroism, which will of course, due to the presence of circularly polarized photons, contribute to the signal in any experiment – whether or not based on structured light. The differential rate of absorption for circularly polarized photons is found to be;

$$\Gamma^{(L(\sigma))} - \Gamma^{(R(\sigma))} = \Delta\Gamma_{CD} + \ell \frac{2\kappa}{r} \varepsilon_{ikm} \hat{k}_m \hat{\phi}_j \mu_k^{f0} Q_{ij}^{f0}, \qquad (12)$$

with

$$\Delta\Gamma_{CD} = \kappa \left[ c^{-1} \left( \delta_{ik} - \hat{k}_i \hat{k}_k \right) \left( \mu_i^{f0} \bar{m}_k'^{f0} - \bar{\mu}_k^{f0} m_i'^{f0} \right) + 2k\varepsilon_{ikm} \hat{k}_m \hat{z}_j \mu_k^{f0} Q_{ij}^{f0} \right], \qquad (13)$$

where $m_i'^{f0} = i m_i^{f0}$. Here, the additional terms in (13) have been derived in a similar method to the well-known results presented by Power and Thirunamachandran [45,50], but with use of the LG-mode expansions (2) and (3). In contrast to the CVD contribution (11) to (12), they exhibit no dependence on the sign of $\ell$.

As it currently stands, the CVD rate (11) is applicable to a system of one or more molecules that are individually fixed in orientation within the beam – or the result may be taken as representative of molecules in a system that possesses a degree of orientational order, such as a poled liquid crystal. If the interaction between a molecule and the local wave vector at one specific point in the beam is of the exact same magnitude but opposite sign to the mirror-symmetric position across the other side of the beam then, for a system of molecules having the *same* orientation relative to one another, the CVD effect will be zero if the rate of absorption is monitored across the whole beam profile (as is usual in experimental studies), with only the standard CD (equation (13)) persisting – see Figure 1. In such a scenario, observation of the CVD effect would be dependent upon probing for locally differential absorption at different locations within the beam. However, if the molecular system possesses only a partial degree of orientational order across the beam, the extent of CVD will be diminished – that is to say, for any single position on one side of the beam, the mirror symmetric position on the other face need not necessarily cancel the CVD effect as the pairs of molecules in question may have slightly different orientations.

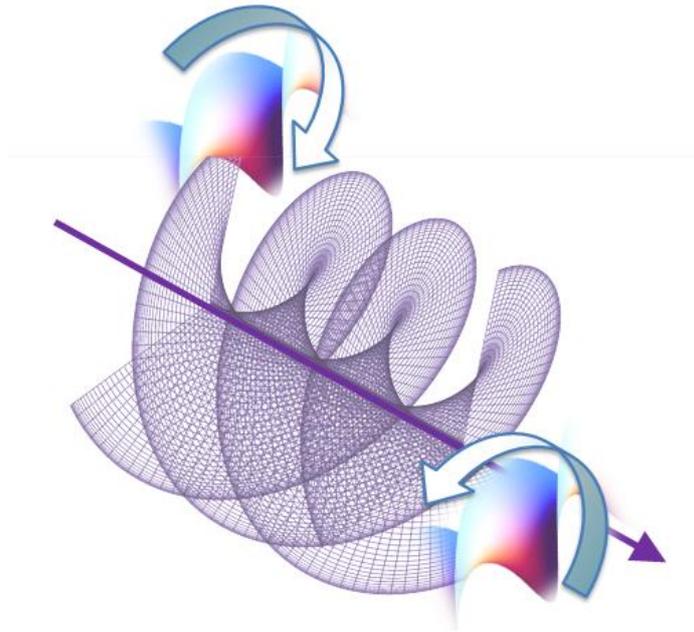

Figure 1: (Color online) Twisted beam, incident from the left, propagating through an anisotropic system of molecules such as a liquid crystal exemplified by two helical structures with a common orientation. Positioned at opposite sides of the beam, each experiences with regard to its own structure a different directional sense of the phase gradient.

To complete the analysis we can now allow for a complete lack of orientational order, as is the case in conventional molecular fluids. To account for this involves implementing an isotropic rotational average on the right-hand side of (11), which brings into effect a contraction of the third rank tensor $\mu_k^{f0} Q_{ij}^{f0}$ with the corresponding third rank isotropic Levi-Civita epsilon [51]. Thus, the random averaging delivers $\langle \mu_k^{f0} Q_{ij}^{f0} \rangle \approx \varepsilon_{ijk} \varepsilon_{\lambda\mu\nu} \langle \mu_\nu^{f0} Q_{\lambda\mu}^{f0} \rangle$ (where angular brackets denote the rotational average). However, since the electric quadrupole moment is symmetric in its indices, and the Levi-Civita tensor is fully index antisymmetric, the molecular average is zero. Therefore, the CVD rate (11) and E1E2 CD contribution in (12) vanishes for randomly oriented molecules. The first term in (13), which represents the contribution from E1M1 coupling does not vanish upon random orientational averaging, represents the well-known contribution to circular dichroism in molecular fluids [3,50] but accounting for the light source being a Laguerre-Gaussian beam. It may therefore be concluded that observing the CVD effect in isotropic molecular fluids would not be possible, and that full verification of the mechanism itself would involve resolving the extent of absorption at local positions in the beam profile of an ordered, or at least partially ordered molecular system.

**IV. ACHIRAL MEDIA**

To tackle the physics of achiral media, it is useful to identify the origins of their distinction from chiral manifestations. The most well-known chiroptical phenomena exhibit discriminatory interactions through an interplay of chiral molecules/media with the handedness of circularly polarized light. This leads to a system possessing two different forms of handedness – one manifest in the symmetry of the material component (e.g. a chiral molecule), and the other depending on whether the electric (and magnetic) field vector circulates in a clockwise or an anticlockwise sense of direction. Here, we have to be careful to distinguish between the criteria for chiroptical effects in randomly ordered or isotropic fluid media, and systems with any degree of orientational order. As we shall see, there is also a difference to be considered with regard to whether the input optical fields are, or are not, structured.

The first difference arises because, for freely rotating molecules, a capacity to exhibit chirality can be correctly interpreted on the basis of a lack of reflection symmetry, which is the standard chemical definition: it is readily shown that reflection, coupled with a $\pi$-rotation about the normal to the reflection plane, has the same effect as spatial inversion. This definition, in itself, would appear to exclude the possibility of isotropic media comprising *achiral* components, possessing at least one element of reflection symmetry (or any axis of improper rotation), exhibiting chiroptically differential effects. However, such a supposition depends on there being no time-independent local field gradients; for irradiation with structured light, it is no longer possible to adopt a criterion that takes no account of such gradients.

Conversely, in systematically *ordered* media, (whether the orientational ordering is total or partial), a lack of full rotational symmetry means that chiroptical effects can be anticipated in either chiral or achiral samples [52–54]. As was observed in Section III, E1E2 terms cannot contribute to chiroptical effects in isotropic chiral or achiral media. Therefore it is the next level of terms, namely E2E2, that we need to focus upon. The magnitude of effects associated with these terms are, it should be noted, generally lower in magnitude than the E1E2 terms that correspond to chiral media.

From (7) the total rate to single-photon absorption by E2E2 contributions is seen to be:

$$\Gamma_{\text{E2E2}} = \kappa e_i \bar{e}_k Q_{ij}^{f0} Q_{kl}^{f0} A_j \bar{A}_l, \qquad (14)$$

where

$$A = \left( \hat{r} \frac{1}{f_{\ell,p}(r)} \partial_r f_{\ell,p}(r) + \frac{1}{r}\left(i\ell\hat{\phi} - \hat{r}\right) + ik\hat{z} \right). \tag{15}$$

On inspection, it is possible to isolate the terms in (14) that will exhibit SOI similar to that highlighted in the CVD effect in Section III:

$$\Gamma'_{E2E2}(\ell) = \kappa \frac{i\ell}{r} e_i \bar{e}_k Q_{ij}^{f0} Q_{kl}^{f0} \left[ \frac{1}{f_{\ell,p}(r)} \partial_r f_{\ell,p}(r)\left(\hat{\varphi}_j \hat{r}_l - \hat{\varphi}_l \hat{r}_j\right) + \frac{1}{r}\left(\hat{\varphi}_l \hat{r}_j - \hat{\varphi}_j \hat{r}_l\right) \right]. \tag{16}$$

As in the previous Section, we utilise identity (10), which allows the rate to be expressed in terms of the helicity eigenvalues of the circularly polarized photons:

$$\Gamma'^{(L/R)}_{E2E2}(\ell) = \sigma\ell \frac{\kappa}{2r} \varepsilon_{ikm} \hat{k}_m Q_{ij}^{f0} Q_{kl}^{f0} \left(\hat{\varphi}_j \hat{r}_l - \hat{\varphi}_l \hat{r}_j\right)\left[ \partial_r \ln f_{\ell,p}(r) - \frac{1}{r} \right]. \tag{17}$$

Once again the result represents a chiroptical effect that is dependent on the product $\sigma\ell$, whose origins lay in SOI. The underlying mechanism for these SOI and in the chiroptical processes highlighted in the previous Section will be discussed in detail in Section V. Distinctly different from the CVD effect, the E2E2 rate (17) is only dependent on the handedness of the radiation through $\sigma\ell$, and not the molecular handedness as $Q_{ij}^{f0}Q_{kl}^{f0}$ is invariant under the spatial parity operation. Therefore, a chiroptical effect can be evinced in achiral matter, solely through the interplay of handedness in the interacting circularly-polarized LG modes of light.

The averaging technique needed to address a system of molecules in completely random orientation has to tackle the fourth-rank tensor in (17), and is therefore a little more involved than the previous case involving third-rank tensor averaging. The procedure can be implemented using standard results [51], noting its local validity for any set of mutually orthogonal axes in three-dimensional space:

$$\left\langle \Gamma'^{(L/R)}_{E2E2}(\ell) \right\rangle = \sigma\ell \left( \frac{2\kappa}{3r} \right)\left( Q_{\lambda\lambda}^{f0} Q_{\pi\pi}^{f0} - Q_{\pi\mu}^{f0} Q_{\mu\pi}^{f0} \right)\left[ \partial_r \ln f_{\ell,p}(r) - \frac{1}{r} \right]. \tag{18}$$

It emerges that the form of this result elicits some interesting new insights into the physics. Analogous to the CVD effect (11), both (17) and (18) are invariant to the transformation $(\sigma,\ell) \to (-\sigma,-\ell)$, but not $(\sigma,\ell) \to (-\sigma,\ell)$ or $(\sigma,\ell) \to (\sigma,-\ell)$. That is to say, for a fixed polarization-handedness, the rate of absorption will be different for a right-handed twist photon than for a left-handed one (the same applies for a fixed helicity of twist and differing handedness of polarization). However, in contrast to the CVD case, the effect (18) does not vanish under a full-rotational average and therefore persists in an isotropic molecular fluid. Equally, since the local 3D average is insensitive to any direction of field gradient, we can conclude that the absorption rate contributions represented by either equation (17) or (18), averaged over the whole cross-section of a structured beam, will be zero.

Before moving on to discuss the remaining terms that exhibit discriminatory effects dependent on the SAM or OAM in (14) (Section VI), we tackle in the next Section the underlying mechanism of the spin-orbit interaction – $\sigma\ell$ – exhibited in the CVD effect (11), along with (17) and (18) for the dichroic-like effect in achiral media.

## V. SPIN-ORBIT INTERACTIONS OF LIGHT

So far we have derived equations to quantify and represent the mathematics of several facets of the chiroptical processes manifested in single-photon absorption. However, the underlying physical mechanism is yet to be fully illuminated, and it is this task we now aim to tackle. It has been explicitly highlighted how the engagement of E2 or higher order transition moments is a requisite if one wishes to observe a chiroptical effect with sensitivity to the handedness of the vortex twist. This, however, is only one part of the overall mechanism at play. As shown above, to secure a real expression for the overall rate (and therefore represent a measurable quantity) requires circularly polarized states to be utilised. This innocuously introduces, into the rate equation, a $\sigma = \pm 1$ term for the helicity eigenvalues of the circularly polarized photons: this in turn engenders the product '$\sigma\ell$'. It is worth dwelling on the nature of such a feature which – in other connections to be discussed below – might be considered to exhibit a form of spin-orbit coupling. At the outset, we should therefore be clear that the sense in which we use the term 'spin-orbit interaction' (SOI) is simply as a marker for a mechanism necessitating the presence of both kinds of optical angular momentum.

In the context of beam optics, changes in the polarization state (and thus the associated helicity) can be engaged in manipulating intensity distributions and propagation directions, producing novel optical phenomena. Importantly, these SOI become distinctly important at the microscopic, sub-wavelength scale: nano-optics, photonics, and the light-matter interactions that take place between photons and molecules, for example. Well-known examples of spin-orbit coupling include the spin- and orbital-Hall effects of light, and spin-to-orbit AM conversion [17]. In the spin-Hall effect [33,55], light experiences spin-dependent position or momentum of light. These arise from coupling between SAM and extrinsic OAM, and a similar vortex-dependent shift known as the orbital-Hall effect occurs between intrinsic and extrinsic OAM [56,57]. The coupling that takes place between the intrinsic SAM and OAM produces the spin-to-orbit AM conversion [58–62]: through a combination of anisotropic media and inhomogeneities the polarization, intensity, and phase distributions can all be manipulated. These SOI effects coupled with surface plasmons have already been utilised in the production of chiroptical interactions sensitive to optical OAM [18,24] and spin-controlled transmission of light [63]. However, in the derivation of the optical rates in this article we have assumed freely-propagating paraxial light, which is neither scattered nor focussed, being absorbed in homogeneous collections of molecules. Therefore, the aforementioned spin-orbit couplings cannot explain the SOI that is taking place in (11) and (17), and to illuminate their origin we must look at the physics of the real electronic transitions taking place during the absorption process.

The well-known plane waves, although technically unphysical, offer exact solutions to Maxwell's equations. These plane waves have zero electric and magnetic field components in the direction of propagation ($\boldsymbol{k}$). The Laguerre-Gaussian modes concentrated on this paper are only *approximate* solutions to Maxwell's full theory [64]. In their application, the paraxial approximation is assumed, whereby the *z*-dependence of the field amplitude is neglected due to the relatively larger transverse variation. It was Lax et al. [65] who highlighted how such an approximation leads to zeroth-order fields which are not exact solutions to Maxwell's equations, though contributing the largest component to the approximate solution, and that the first-, second- and higher order corrections to these paraxial solutions offer small corrections through an expansion parameter $(kw_0)^{-1}$.

The zeroth-order solutions represent completely transverse fields whereas, due to the divergence-free character of the fields, the small first order component must be longitudinal.

The mode expansions (2) and (3) have been derived within the paraxial approximation [43,44], and as can be seen from the interaction Hamiltonian (1), both the E1 and M1 couplings are clearly interacting with the zero-order transverse fields. However, the E2 coupling that is dependent on both the transverse and longitudinal field gradient is interacting with higher-order fields. This is because the transverse gradient of the zero-order field is clearly engaged in twisted light interactions with E2 moments, through the $\partial_r$ and $\partial_\phi$ operations in (6). Of course, we have already seen that the most important feature is that the $\partial_\phi$ operation leads to the unique dependence on the topological charge, which in turn emerges in the form of an SOI through $\sigma\ell$.

In the equations that have exhibited discriminatory photon absorption through SOI, it is important to note that all have been dependent on the transverse gradient of the zero-order field, and as such can be seen to be interacting with longitudinal field components: the E1E2 rate (11) came from a $\partial_\phi$-dependent term; the E2E2 rate (17) from a $\partial_\phi \partial_r$-dependent term. This explains why, in previous work [14] restricted to the dipole interactions E1 and M1, no role was found for the topological charge of a vortex beam in chiroptical interactions: the underlying approximation did not allow for longitudinal components of the field, and thus precluded SOI phenomena. Interestingly, beams with a longitudinal component have been shown to produce a similar discriminatory effect in both achiral and chiral media [66,67] and play an important role in twisted-light-matter interactions, particularly for beams with antiparallel OAM and SAM [68].

It is worth highlighting a subtle point about the SOI optical phenomena studied so far, and the relationship between paraxial and non-paraxial solutions to the wave equation. It appears that adopting the more generally applicable non-paraxial form for an LG mode would lead to no new, additional physics than already highlighted using the paraxial form as we have done for single-photon absorption. That is to say, using a non-paraxial LG beam in the derivations presented would not introduce any additional features, beyond those already extracted; their effect would only be to introduce terms with marginal quantitative effect. This is due to the fact that the azimuthal and radial phase factor is left unchanged when transitioning from the paraxial to the non-paraxial solution; as has been clearly shown above, it is this factor which produces chiroptical effects with regards to the topological charge. In the construction of a non-paraxial form of a LG beam, satisfaction of the complete Helmholtz equation is obtained by simply changing the longitudinal phase factor only in the paraxial form [69,70]. This is why the results presented by Afanasev et al. [19] 'remain large in a paraxial limit'. It is worth emphasising this point: in studying interactions of twisted light that induce electronic transitions in matter during absorption, neither paraxial or non-paraxial beams of light can produce SOI in optical processes that involve only dipole (E1 and M1) coupling – precisely because they don't involve quadrupole E2 transition moments. These E2 transitions are important because they involve the transverse and longitudinal gradient of the field, the transverse part of which involves longitudinal electric field components coupling to the matter. Terms that depend on $\partial_z$ will differ between paraxial and non-paraxial representations, but the extent of that change should be of only qualitative significance.

Another important piece of the puzzle is the role of angular momentum transfer that takes place in electronic transitions when photons are absorbed by atoms and molecules. Just as the intrinsic SAM of a photon is known to be transferred to the orbital angular momentum of an electron, i.e. the internal degrees of freedom of an atom/molecule, the question has naturally been asked whether the OAM of a twisted light beam might be transferred to the electronic motion. Different theoretical studies provided conflicting conclusions [71–75]: however, it became widely agreed that in dipole transitions, any OAM is transferred to external degrees of freedom – i.e. those associated with movement of the atom or molecule as a whole. However, early theoretical predictions [71] that optical OAM might indeed be transferred to the internal degrees of freedom through electric *quadrupole* transitions have been vindicated by recent experimental evidence [76]. In any such E2 transition, the intrinsic single unit of SAM of a photon is transferred to the electronic motion, along with a single unit of OAM from the beam, with the remaining $(\ell \pm 1)\hbar$ OAM transferred to the center of mass motion. To be clear, linearly polarized photons can induce an electronic quadrupole transition, and so can circularly polarized photons: it cannot generally be assumed that there is a one-to-one correlation between a transition multipole involved in absorption, and the angular momentum content of the incident field [77]. It is an extension of the same principle that applies to photons possessing OAM. Selection rules are based on changes in angular momentum of the molecule, and are not solely constricted to the addition of angular momentum from a photon.

Therefore, due to the fact that in E2 transitions both SAM and OAM are transferred to the electronic internal degrees of freedom of the molecule, rather than the latter being conveyed to the center of mass motion as happens in dipole transitions, the intrinsic nature of the OAM of paraxial beams can in fact be registered on a local scale. The SOI that provide a basis for the chiroptical effects in molecules discussed in the paper so far are therefore distinctively different from those occurring in non-paraxial beams of light that have been focused or scattered, or those occurring in plasmonic-enhanced vortex light-matter interactions. As regards electric multipoles, both the SAM and OAM of light must participate in transitions that are simultaneously electric-dipole and electric quadrupole-allowed, in order to observe these local chiroptical effects in chiral molecules, or solely electric quadrupole-allowed in achiral media.

## VI. 2D CHIRALITY

In the previous sections we have highlighted certain kinds of chiroptical effect that can occur with twisted light in its interactions with both chiral and achiral molecules. These interactions, sensitive to the sense of twist and the circular polarization of the input beam, have been shown to owe their primary origins to the SOI that occur in a process entailing an E2 transition moment. However, calculating the total rate for the E2E2 contributions (14) also produces terms that show a chiroptical sensitivity to the wavefront twist *or* circularly polarization state, and hence the OAM *or* the SAM of light, but which are not SOI: it is these terms we shall concentrate upon in this Section.

It transpires that a key feature of these terms, neglected in previous treatments, is their manifestation in systems with two-dimensional (2D) chirality [78–80]. The associated symmetry rules are distinctively different from the 3D form, though equally strict. A surface or other locally ordered structure feature may be achiral with regard to one specific orientation of beam incidence, yet exhibit chirality with respect to another. Indeed, the reason for the connection of circular polarization with optical helicity is the propagating character of light: the electric and magnetic vectors sweep out a helical locus, rather than the circle that would otherwise arise. When circularly polarized light encounters an orientationally fixed component, however, sensitivity to its helicity demands satisfaction of only the conditions for 2D chirality, for which it is *sufficient* to break reflection symmetry in one plane containing the propagation vector. Indeed, much of the recent research on chiral effects involving metasurfaces is concerned with photonic interactions between optical beams and sculpted material structures of just such a 2D chiral form [81]. Gammadion surface features, which have $C_4$ symmetry within the surface plane, have become a common motif in such studies [82]; however, the presence of one plane of reflection symmetry may still allow the exhibition of chiroptical behaviour [83].

In the present connection, it becomes evident that 2D chirality can be exhibited with respect to incident light with circular polarizations and/or a twisted wavefront. The upshot is that certain terms in the generic single-photon rate equation, (7), that cannot support chiroptical effects in fluid media, may nonetheless do so in ordered media, by exploiting 2D chirality. The initial contributions with this capacity are in fact the first terms on the right-hand side of (14): these deserve special attention, for they illustrate the possibility of a dichroic effect that may occur with linearly polarized light. The two terms may together be cast as $\partial_r \ln f_{\ell,p}(r)$, immediately signifying that for *un*structured light – a traditional plane wave for example – the absence of any dependence on distance from the beam center (or indeed any axis) will mean that this E1E2 term in its entirety will deliver a rate zero contribution. For the structured forms of light that are our focus, however, this is not the case.

Moreover, the field components signified by the indices *i*, *j* and *k* are necessarily coplanar – in the paraxial approximation. However, whereas the *i* and *k* components of a plane polarized electric field might legitimately be referred to a standard Cartesian basis in the plane perpendicular to $\hat{z}$, (the $\hat{x}$ and $\hat{y}$ directions in a laboratory-fixed frame of reference, for example, identifiable from 2D rotation of the $\hat{r}$ and $\hat{\phi}$ directions at any chosen off-axis location and then applied across the whole beam), the *j* component specifically refers to a unit vector in the radial direction within that same plane. The key distinction is that the latter is a necessarily position-dependent quantity, its sign dependent on 'side' of the beam; clearly this does not apply to the *i* and *k* field components. Thus, a molecule with 2D (or 3D) chirality positioned on one side of the beam will generate an E1E2 rate contribution that is opposite in sign from a diametrically opposed molecule of the same conformation. Equally, a molecule of space-inverted symmetry at the same original position will also generate a rate contribution of the opposite sign. 3D chirality is not required: 2D chirality suffices. In fact, in both of these instances 2D chirality signifies essentially the same effect: placing a mirror across the center of the beam, along rather than transverse to the beam axis, has the same effect of inverting 2D molecular symmetry as inverting the sense of the radial unit vector. These considerations now elucidate the analysis that follows.

Evidently, as the total E2E2 rate (14) depends quadratically on ***A*** (15), which itself contains 4 terms, there are in total 16 contributions to the E2E2 rate of single-photon absorption. It is important to recognize that 8 of these 16 terms contribute to the rate of absorption without any chiroptical significance – they designate the appropriate contribution to E2E2 from an LG beam just as they would for a plane-wave light beam. These terms are in fact easily identified by being 'quadratically' dependent on components of a single unit-vector, i.e. $\hat{r}_i \hat{r}_j$, $\hat{\phi}_i \hat{\phi}_j$, and $\hat{z}_i \hat{z}_j$. The new and more interesting physics arises from cross-terms, one of which has already been explicitly highlighted in previous Section through (17), which is of course a cross-term dependent on $(\hat{\phi}, \hat{r})$. A further pair of cross-terms which involves $(\hat{z}, \hat{r})$, since it carries a factor of *i*, vanishes for plane polarizations: for circular polarizations it may be expressed as:

$$\Gamma'^{(L/R)}_{\text{E2E2}}(\hat{z},\hat{r}) = \sigma \frac{\kappa}{2} k \varepsilon_{ikm} \hat{k}_m Q^{f0}_{ij} Q^{f0}_{kl} \left( \hat{z}_j \hat{r}_l - \hat{z}_l \hat{r}_j \right) \left( \partial_r \ln f_{\ell,p}(r) - \frac{1}{r} \right), \tag{19}$$

This term depends on the helicity of light, i.e. the handedness of circular polarization (there is a dependence on topological charge through $\partial_r \ln f_{\ell,p}(r)$, but that involves only the modulus, $|\ell|$). The last remaining pairs of cross-terms engage $(\hat{z}, \hat{\phi})$, and their contribution depends on the sign of the topological charge:

$$\Gamma'_{\text{E2E2}}(\hat{z},\hat{\phi}) = \ell \frac{k}{r} \kappa e_i \bar{e}_k Q^{f0}_{ij} Q^{f0}_{kl} \left( \hat{\phi}_j \hat{z}_l + \hat{z}_j \hat{\phi}_l \right), \tag{20}$$

notably, this term persists even with linearly polarized light. Indeed, similar chiroptical effects that show discriminatory behaviour dependent on the sign of $\ell$ for linearly polarized twisted beams have been previously reported [21]. It is worth emphasizing that the 2D chiroptical phenomena detailed by (19) and (20) both entail a dependence on the material symmetry through the orientation-dependent products of E2 molecular transition moments. This marks a distinction from the discriminatory effects that occur in chiral plasmonics [84,85].

Before concluding, it is briefly worth noting that the contribution of M1E2 terms to the optical absorption rate will generally be of the same order of magnitude to the E2E2 terms, and simple extraction of the appropriate terms from (7) allows their rate contribution to be secured. Notably, since both the magnetic dipole and electric quadrupole operators are of even parity, this rate contribution can be present irrespective of the molecular symmetry, provided the absorption transition is allowed by both forms of multipole (as determined by the irreducible representations of the initial and final state, in the point group for the appropriate molecular symmetry). Accordingly, no additional features of compelling interest arise from this term, and we consider it no further.

## VII. CONCLUSION

This aim of this work has been to explicitly highlight the major role that optical orbital angular momentum can play in chiroptical light-matter processes. Specifically, we have concentrated on single-photon absorption in both chiral and achiral media. It was shown how in conventional 3D materials such effects are manifest through SOI, and that in a system of chiral molecules there must be a degree of orientational order to observe the CVD effect; conversely, for achiral molecules the dichroic-like absorption persists even when the system has full rotational symmetry. Alongside these chiroptical interactions, manifest through SOI, we also identified further phenomena which exists as a result of the optical OAM and SAM interacting in a discriminatory fashion with systems possessing 2D chirality. Once again, the unparalleled utility of quantum electrodynamics for studying fundamental light-matter interactions has been exhibited: even for the simplest of optical processes, single-photon absorption, a multitude of new physical phenomena have been extracted and quantified.

Summarising the key findings in the general case of both chiral and achiral molecules: chiroptical interactions with twisted light necessitate the engagement of electric quadrupole (or higher) transition moments as they depend on the transverse field gradient which produce longitudinal field components – the cause of the SOI. The origins of optical SOI are intrinsic in Maxwell's equations, the foundation of QED. Therefore, it is of no surprise that the theory of QED accounts for such interactions, and they fall naturally out of the theory when suitably applied. It proves to be of little significance whether the light is paraxial or non-paraxial, in the case of single-photon absorption by molecules from freely-propagating beams; the important factor is the E2 transition moment, which allows for intrinsic SAM and OAM be registered in a real electronic transition and thus engender SOI. In both chiral and achiral media, chiroptical effects are sensitive to both the helicity of light $\sigma$ and the topological charge $\ell$ through the product '$\sigma\ell$'. In contrast to these SOI, further phenomena in molecular systems possessing 2D chirality are seen to be sensitive to $\sigma$ $or$ $\ell$, but not both simultaneously.

We believe it clearly evident that the field of chiroptical interactions engaging the OAM of light has in recent years began to flourish: with theoretical developments of the nature highlighted in this paper, the true potential of the area will be increasingly realised over the coming years. Our future efforts will concentrate on scattering and more exotic multiphoton optical processes. Indeed, studies looking at differential Raman [86] and Mie [87,88] scattering suggest that the sign of $\ell$ introduces a sensitivity to these scattering processes, and in a recent study it has even emerged that there exists an optical OAM-sensitive transmission rate of LG beams through mouse brain tissue [89].